\documentclass[journal]{IEEEtran}

\makeatletter
\long\def\@makecaption#1#2{\ifx\@captype\@IEEEtablestring%
\footnotesize\begin{center}{\normalfont\footnotesize #1}\\
{\normalfont\footnotesize\scshape #2}\end{center}%
\@IEEEtablecaptionsepspace
\else
\@IEEEfigurecaptionsepspace
\setbox\@tempboxa\hbox{\normalfont\footnotesize {#1.}~~ #2}%
\ifdim \wd\@tempboxa >\hsize%
\setbox\@tempboxa\hbox{\normalfont\footnotesize {#1.}~~ }%
\parbox[t]{\hsize}{\normalfont\footnotesize \noindent\unhbox\@tempboxa#2}%
\else
\hbox to\hsize{\normalfont\footnotesize\hfil\box\@tempboxa\hfil}\fi\fi}
\makeatother

\usepackage{graphicx}
\usepackage{amssymb,amsmath,bm}
\usepackage{multirow}
\usepackage{cite}
\usepackage{hhline}
\usepackage{booktabs}

\begin{document}

\title{Speaker Separation Using Speaker Inventories and Estimated Speech}

\author{Peidong~Wang,
        Zhuo~Chen,
        DeLiang~Wang,~\IEEEmembership{Fellow,~IEEE,}
        Jinyu~Li,
        and~Yifan~Gong
        \thanks{This research started during P. Wang's internship at Microsoft and finished at The Ohio State University.}
        \thanks{P. Wang is with the Department of Computer Science and Engineering, The Ohio State University, Columbus, OH 43210 USA (e-mail: wang.7642@osu.edu).}
        \thanks{Z. Chen is with Microsoft, Redmond, WA 98052 USA (e-mail: zhuc@microsoft.com).}
        \thanks{D. L. Wang is with the Department of Computer Science and Engineering and the Center for Cognitive and Brain Sciences, The Ohio State University, Columbus, OH 43210 USA (e-mail: dwang@cse.ohio-state.edu).}
        \thanks{J. Li is with Microsoft, Redmond, WA 98052 USA (e-mail: jinyli@microsoft.com).}
        \thanks{Y. Gong is with Microsoft, Redmond, WA 98052 USA (e-mail: \newline yifan.gong@microsoft.com).}
}
\maketitle

\begin{abstract}
We propose speaker separation using speaker inventories and estimated speech (SSUSIES), a framework leveraging speaker profiles and estimated speech for speaker separation. SSUSIES contains two methods, speaker separation using speaker inventories (SSUSI) and speaker separation using estimated speech (SSUES). SSUSI performs speaker separation with the help of speaker inventory. By combining the advantages of permutation invariant training (PIT) and speech extraction, SSUSI significantly outperforms conventional approaches. SSUES is a widely applicable technique that can substantially improve speaker separation performance using the output of first-pass separation. We evaluate the models on both speaker separation and speech recognition metrics.
\end{abstract}

\IEEEpeerreviewmaketitle

\section{Introduction}
\label{sec:intro}



\IEEEPARstart{S}{peech} overlaps occur commonly in daily conversations. They make automatic speech recognition (ASR) and speaker diarization in conversations difficult. The task of separating overlapped speech is referred to as speaker (or speech) separation and has long been an active research area.

A key challenge in speaker separation is the so-called permutation problem as defined in \cite{hershey2016deep}. When multiple speakers are involved in a speech mixture, different orders of output signals may lead to conflicting gradients across training utterances. Two kinds of algorithms were proposed to handle the permutation problem, namely speaker separation and speech extraction. Speaker separation uses specifically designed training objectives that are invariant to the order of the outputs. Deep clustering \cite{hershey2016deep,isik2016single} and permutation invariant training (PIT) \cite{yu2017permutation,kolbaek2017multitalker} are two representative approaches. Many studies have been conducted to improve these two approaches, including new objective functions \cite{wang2018alternative,luo2019augmented,chen2017deep}, end-to-end training \cite{xu2019optimization,chen2018sequence,luo2018speaker,li2019listening,li2018source}, new model architectures \cite{li2018deep,Liu2019DivideSeparation,xu2018single}, and different input features \cite{luo2019conv,wu2019time,wang2018multi,yoshioka2018multi,gu2019end,shi2019furcanext}. Speech extraction avoids the permutation problem by extracting only one output signal using the bias information distinguishing the target speaker from others. The bias information can be in the form of visual signals \cite{ephrat2018looking,zhao2018sound,wu2019time}, speaker locations \cite{chen2018location,perotin2018multichannel,zhao2018two}, and speaker identities (SIDs) \cite{zmolikova2017speaker,delcroix2018single,wang2018voicefilter,Wang2018DeepMixtures,xiao2019single,ochiai2019unified}. Among these three types of bias information, SIDs are easier to acquire since they do not need extra hardware such as cameras and microphone arrays. Speaker identities are readily available in many scenarios such as meetings. For SID based speech extraction, Delcroix \emph{et al.} proposed a method called SpeakerBeam to adapt sub-layers in a context-adaptive deep neural network to a target speaker \cite{zmolikova2017speaker,delcroix2018single}. The VoiceFilter proposed by Wang \emph{et al.} \cite{wang2018voicefilter} concatenates spectral features with a d-vector generated by an SID model to extract the speech of the target speaker. Wang \emph{et al.} performed speech extraction using a deep extractor network (DENet) \cite{Wang2018DeepMixtures} formed by stacking two deep attractor networks (DANets) \cite{chen2017deep}. The output of an ``anchor'' (i.e. speaker profile) based DANet is used as input features to another DANet. Xiao \emph{et al.} proposed an attention based speech extraction model \cite{xiao2019single}, which uses an attention mechanism to generate context-dependent biases for speech extraction. Recently, Ochiai \emph{et al.} proposed ASENet, a unified framework for speaker separation and extraction \cite{ochiai2019unified}. They use an attention mechanism to combine the internal embedding vectors of overlapped speech and the embedding of the target speaker profile. Both speaker separation and speech extraction have limitations. Although speaker separation can be used in cases when speaker profiles are not available, they cannot obtain very high separation performance due to the lack of ability to leverage speaker information. Since speech extraction can only generate one output signal, its computation cost would be proportional to the total number of speakers in a meeting; even if a speaker does not say anything in the whole meeting, one would need to launch a speech extraction model for the speaker. Also, speech extraction is performed without the awareness of competing speakers, which may result in insufficient discrimination between some speaker pairs.

We thus propose SSUSI to deal with the issues in both speaker separation and speech extraction. SSUSI leverages bias information to improve separation performance and generates all separated signals in overlapped speech simultaneously. It works equally well or better than speaker separation when some speaker profiles are missing; in such cases, speech extraction is not able to function.

\begin{figure*}[htb]
\centering
\centerline{\includegraphics[width=.8\linewidth]{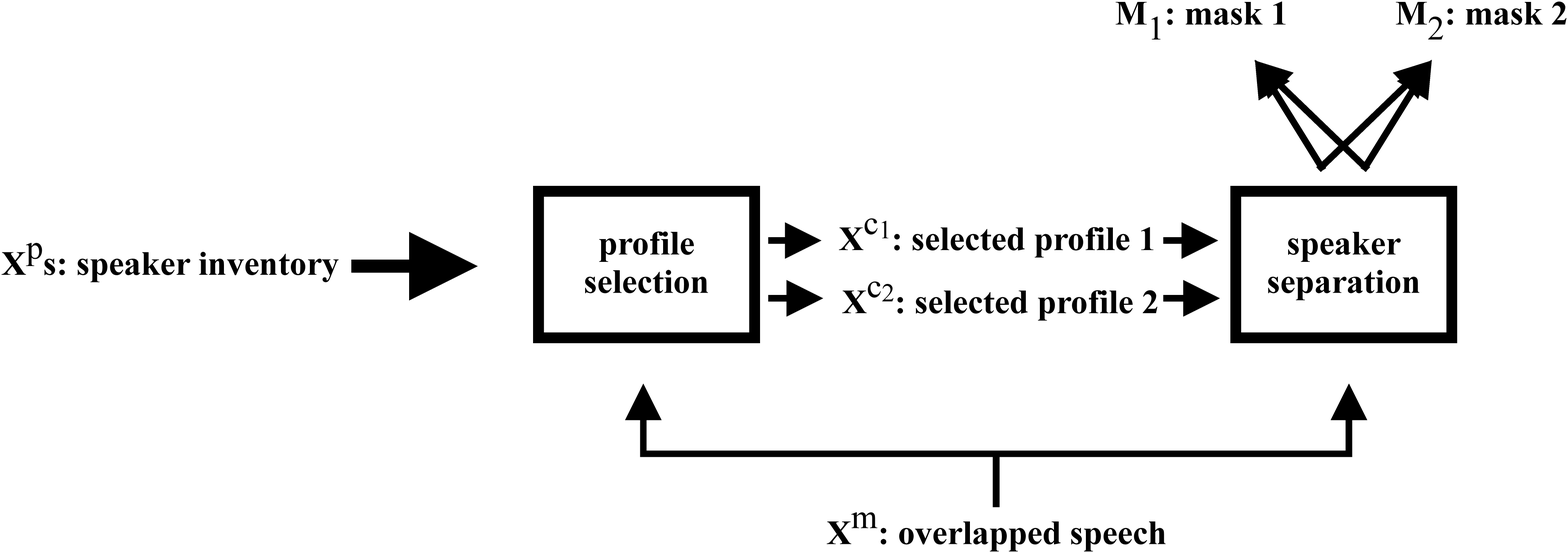}}
\caption{Overview of SSUSI. The features of selected speaker profiles are denoted as $\mathbf{X}^{c_1}$ and $\mathbf{X}^{c_2}$. The arrows between the speaker separation stage and estimated speech denote PIT.}
\label{fig:ssusi}
\end{figure*}

Although SSUSI improves speaker separation in cases when speaker profiles are available, it has two limitations. First, when the number of speaker profiles increases, it is more likely for SSUSI to select wrong speaker profiles and the performance of SSUSI degrades accordingly. Second, when speaker profiles are not available, SSUSI reduces to a normal PIT-based system and its separation performance is relatively low. We thus propose SSUES to deal with these limitations. SSUES takes the estimated speech of first-pass separation as the bias information for another iteration of speaker separation. Since estimated speech is guaranteed to be from a speaker in overlapped speech, the wrong profile selection problem is alleviated. SSUES shows substantial improvement when used with not only SSUSI but also PIT-based first-pass separation. This suggests the wide applicability of SSUES. There are few prior studies on using estimated speech to improve speaker separation performance. Hu and Wang proposed to alleviate the signal level mismatch problem between training and test by adapting a Gaussian mixture model based speaker separation model with estimated signal-to-noise ratios (SNRs) \cite{Hu2013AnSeparation}.

Preliminary results of this paper are presented in \cite{Wang2019SpeechInventory}. This paper expands \cite{Wang2019SpeechInventory} in the following ways. First, we propose SSUES, which makes SSUSI robust to increasing meeting participants. SSUES can also improve the performance of conventional speaker separation methods. Second, we integrate SSUSI and SSUES into SSUSIES, a new speaker separation approach different from existing speaker separation and speech extraction methods in its ability to leverage bias information in a multi-output separation framework.

The rest of this paper is organized as follows. We describe SSUSI and SSUES in Sections \ref{sec:ssusi} and \ref{sec:ssues}, respectively. Experimental setup and evaluation results are presented in Section \ref{sec:exp} and \ref{sec:result}. Concluding remarks are given in Section \ref{sec:conclusion}.

\section{Speaker Separation Using Speaker Inventories}
\label{sec:ssusi}
A speaker inventory consists of a list of speaker profiles collected from the speakers that are possibly involved in overlapped speech. We denote the speakers in the speaker inventory as candidate speakers, and those that are actually involved in overlapped speech as relevant speakers at a certain time. In a scheduled business meeting scenario, for example, speaker profiles can be the prior voice recordings from all meeting invitees. In this paper, the number of relevant speakers is assumed to be two and the speaker inventory only contains voice recordings.

An overview of SSUSI is shown in Fig. \ref{fig:ssusi}. SSUSI performs speaker separation in two stages. First, it selects relevant speaker profiles from candidate profiles using an attention mechanism measuring the correlations between overlapped speech and speaker profiles. After that, two selected profiles are incorporated in the speaker separation stage by a different attention mechanism. The speaker separation stage is designed to exploit the speaker information for separation.


\begin{figure*}[htb]
\begin{minipage}[b]{.42\linewidth}
  \centering
  \centerline{\includegraphics[width=.94\linewidth]{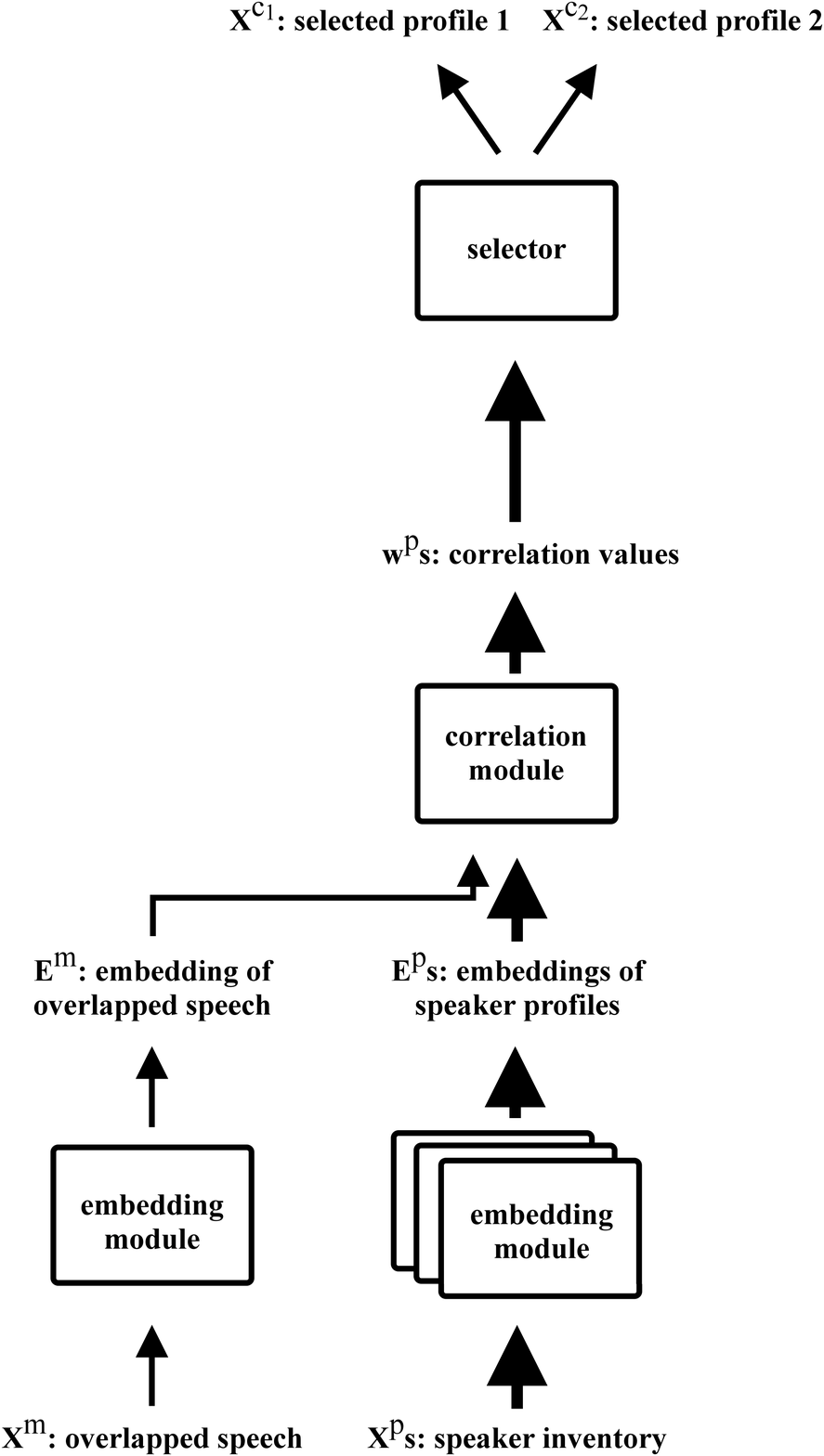}}
  \centerline{(a) profile selection stage}\medskip
\end{minipage}
\hfill
\begin{minipage}[b]{.58\linewidth}
  \centering
  \centerline{\includegraphics[width=\linewidth]{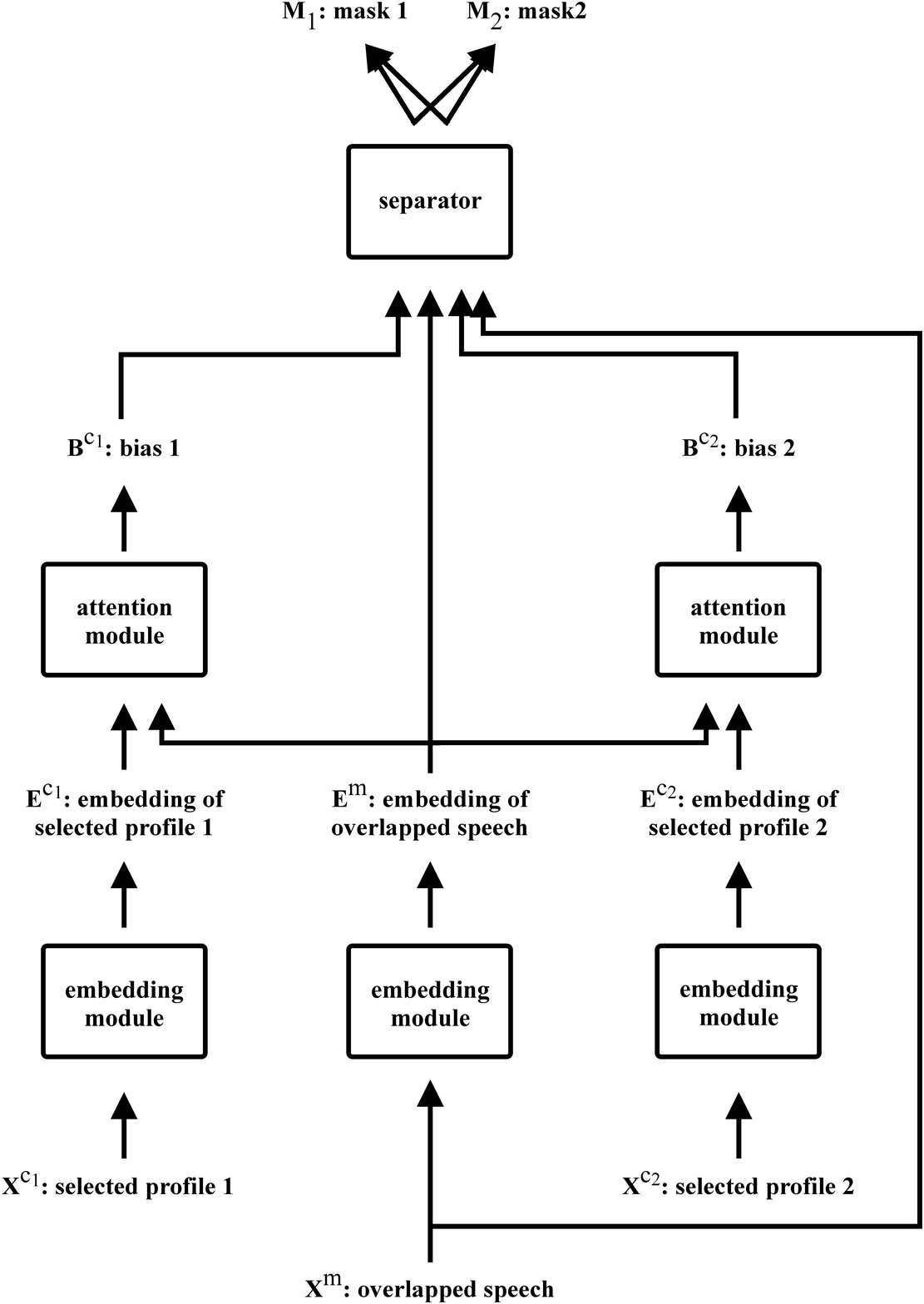}}
  \centerline{(b) speaker separation stage}\medskip
\end{minipage}
\caption{Illustrations of profile selection and speaker separation in SSUSI. The embeddings of selected profiles are denoted as $\mathbf{E}^{c_1}$ and $\mathbf{E}^{c_2}$, and the corresponding speaker biases are $\mathbf{B}^{c_1}$ and $\mathbf{B}^{c_2}$.}
\label{fig:details}
\end{figure*}

\subsection{Profile Selection Stage}
\label{ssec:ssusi_profile}
This stage consists of three components, an embedding module, a correlation module, and a profile selector. Fig. \ref{fig:details}(a) depicts the profile selection stage in SSUSI.

We use a learnable embedding module to extract features for correlation. The embedding module maps input features $\mathbf{X} \in \mathbb{R}^{T \times F}$ to embeddings $\mathbf{E} \in \mathbb{R}^{T \times E}$, where $T$ denotes the number of frames, $F$ refers to input feature dimension, and $E$ is the embedding dimension. For overlapped speech $\mathbf{X}^m \in \mathbb{R}^{T_m \times F}$, the embedding can be denoted as $\mathbf{E}^m \in \mathbb{R}^{T_m \times E}$. For a profile $p$ in speaker inventory $\mathbf{P}$, the embedding can be written as $\mathbf{E}^p \in \mathbb{R}^{T_p \times E}$. Here $T_m$ and $T_p$ denote the numbers of frames in overlapped speech and speaker profile $p$, respectively.

The correlation module measures the correlation between the embedding of overlapped speech and that of each speaker profile. We use $\boldsymbol{e}^m_i$ to denote the vector in $\mathbf{E}^m$ at time frame $i$ and $\boldsymbol{e}^p_j$ the vector in profile embedding $\mathbf{E}^p$ at frame $j$, with $i$ ranging from $1$ to $T_m$ and $j$ from $1$ to $T_p$. We perform correlation in three steps. First, for each profile $p$, we calculate the dot product between each $\boldsymbol{e}^m_i$ and $\boldsymbol{e}^p_j$. Second, we normalize the dot products using the softmax function below. Finally, we average the correlation values over both $i$ and $j$. These three steps are expressed as:
\begin{equation}
    \label{eq:d}
    d^p_{i,j} = \boldsymbol{e}^m_{i} \cdot \boldsymbol{e}^p_{j}
\end{equation}
\begin{equation}
    \label{eq:correlation}
    w^p_{i,j} = \frac{\exp(d^p_{i,j})}{\sum_{p \in \mathbf{P}} \sum_{j=1}^{T_p} \exp(d^p_{i,j})}
\end{equation}
\begin{equation}
    \label{eq:average}
    w^p = \frac{\sum_{i=1}^{T_m} \sum_{j=1}^{T_p} w^p_{i,j}}{T_m T_p}
\end{equation}
where symbol $\cdot$ denotes the dot product operation and $d^p_{i,j}$ is the dot product of embedding vectors $\boldsymbol{e}^m_{i}$ and $\boldsymbol{e}^p_j$. Note that the denominator in equation (\ref{eq:correlation}) is a summation over both profile time steps $j$ and profiles $p$. Symbol $w^p$ is the mean correlation value for speaker profile $p$. The higher $w^p$ is, the more likely that the speaker corresponding to $p$ is involved in the overlapped speech.

The profile selector then selects the first and second largest $w^p$. We denote the selected two profiles as $c_1$ and $c_2$. The profile selection functions are:
\begin{equation}
    \label{eq:argtop1}
    c_1 = \underset{p \in \mathbf{P}}{\arg\max} \{w^p\}
\end{equation}
\begin{equation}
    \label{eq:argtop2}
    c_2 = \underset{p \in \mathbf{P}-\{c_1\}}{\arg\max} \{w^p\}
\end{equation}

\begin{figure*}[htb]
\centering
\centerline{\includegraphics[width=.8\linewidth]{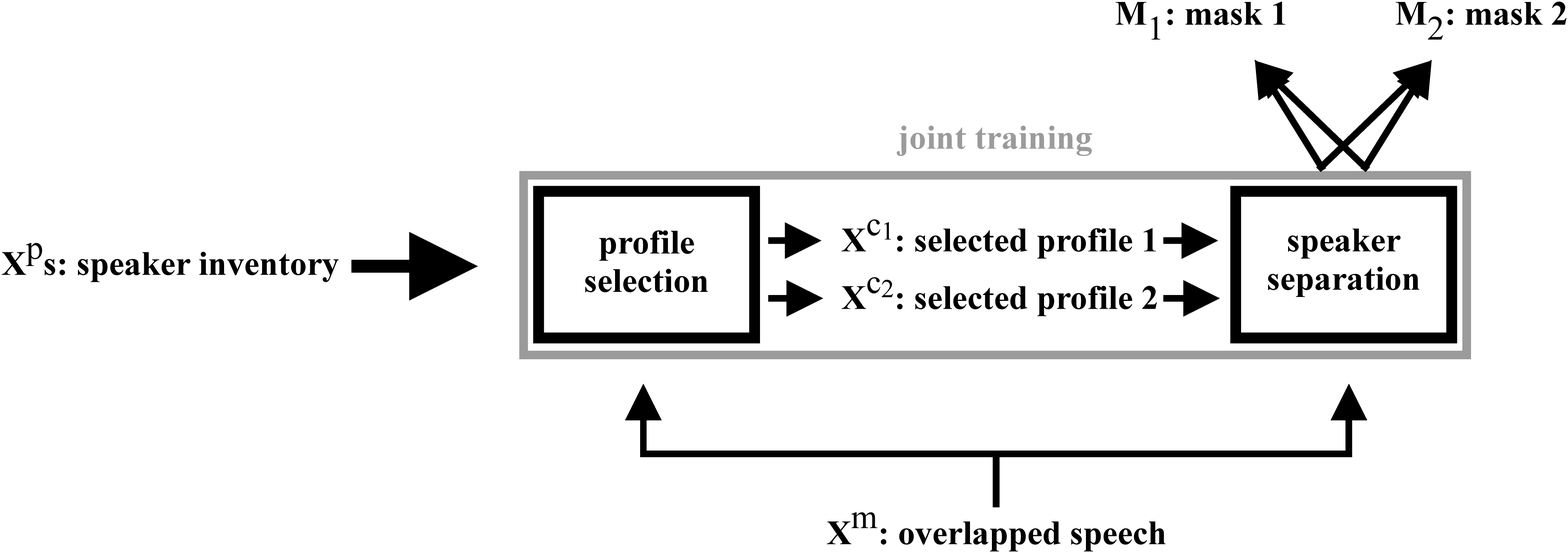}}
\caption{Illustration of SSUSI-JT. The gray box and text indicate the joint training of the profile selection stage and the speaker separation stage.}
\label{fig:ssusi_jt}
\end{figure*}

\begin{figure*}[htb]
\centering
\centerline{\includegraphics[width=.64\linewidth]{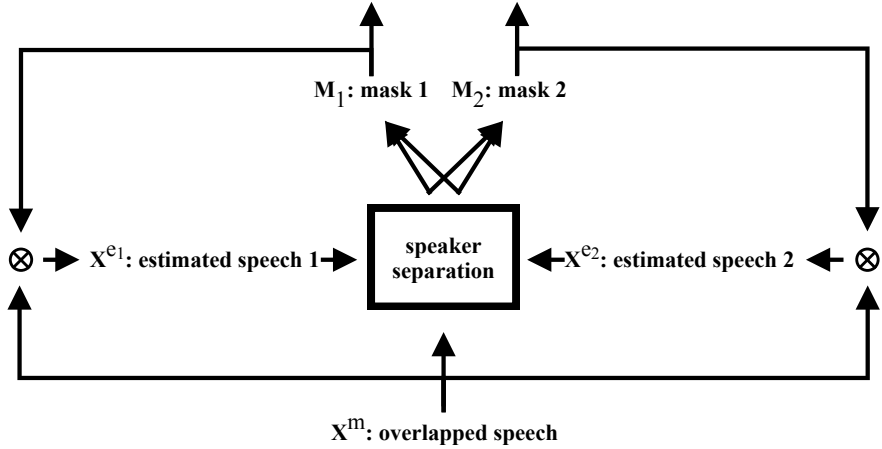}}
\caption{Illustration of SSUES.}
\label{fig:ssues}
\end{figure*}

\subsection{Speaker Separation Stage}
\label{ssec:ssusi_speech_sep}
This stage has three components, an embedding module, an attention module, and a separator. Fig. \ref{fig:details}(b) shows the speaker separation stage in the SSUSI framework.

Similar to the profile selection stage, the embedding module in the speaker separation stage maps input features to embeddings for subsequent attention calculation. For this module, we can re-use the one in the profile selection stage or train a new one specifically for speaker separation, as will be discussed in Section \ref{ssec:ssusi_pse_jt}.

The attention module in the speaker separation stage is slightly different from the correlation module in the profile selection stage. It is used to softly align speaker profiles so that they have the same length as overlapped speech. We denote the aligned speaker profiles as speaker biases since they bias speaker separation towards selected speakers. Speaker bias $\boldsymbol{b}^{c_1}_{i}$ for selected profile $c_1$ at time $i$ is calculated by the following equations:
\begin{equation}
    \label{eq:s_dot}
    d^{c_1}_{i,j} = \boldsymbol{e}^m_{i} \cdot \boldsymbol{e}^{c_1}_{j}
\end{equation}
\begin{equation}
    \label{eq:s_correlation}
    \alpha^{c_1}_{i,j} = \frac{\exp(d^{c_1}_{i,j})}{\sum_{j=1}^{T_{c_1}} \exp(d^{c_1}_{i,j})}
\end{equation}
\begin{equation}
    \label{eq:s_context}
    \boldsymbol{b}^{c_1}_{i} = \sum_{j=1}^{T_{c_1}} \alpha^{c_1}_{i,j} \boldsymbol{e}^{c_1}_{j}
\end{equation}
where $\alpha^{c_1}_{i,j}$ denotes element (${i,j}$) of the attention matrix. Speaker bias $\boldsymbol{b}^{c_2}_{i}$ is calculated similarly. Note that equation (\ref{eq:s_dot}) is the same as equation (\ref{eq:d}) for profile selection. Attention matrix element $\alpha^{c_1}_{i,j}$ in equation (\ref{eq:s_correlation}) differs from correlation matrix element $w^p_{i,j}$ in equation (\ref{eq:correlation}) in that $\alpha^{c_1}_{i,j}$ is normalized over selected profile $c_1$, whereas $w^p_{i,j}$ is normalized over all the profiles in the speaker inventory. Because of this difference, $\alpha^{c_1}_{i,j}$ is able to softly align the embeddings of the selected profiles, whereas $w^p_{i,j}$ is used for comparisons between different profiles.

The separator takes as input the original input features of overlapped speech, the embedding of overlapped speech, and the speaker biases generated from the attention module. The output of the separator are time-frequency masks $\mathbf{M}_1$ and $\mathbf{M}_2$. The training objective is to minimize a signal restoration loss \cite{weninger2014discriminatively,erdogan2015phase} based on PIT. Let $\mathbf{Y}_1$ and $\mathbf{Y}_2$ be the target clean features. An utterance-wise PIT loss can be expressed as equations (\ref{eq:pit}) and (\ref{eq:pit_l}) below:
\begin{equation}
    \label{eq:pit}
    L(\theta) = \min \{l_{1,1} + l_{2,2}, l_{1,2} + l_{2,1}\}
\end{equation}
where $L$ denotes the loss of a training sample and $\theta$ refers to learnable parameters. $l_{u,v}$ means the loss between estimated and clean speech, which is defined as:
\begin{equation}
    \label{eq:pit_l}
    l_{u,v} = || \mathbf{M}_u \otimes \mathbf{X}^m - \mathbf{Y}_v ||_F^2,
\end{equation}
where $||\cdot||_F$ denotes matrix Frobenius norm and $\otimes$ is the element-wise multiplication.


\begin{figure*}[htb]
\centering
\centerline{\includegraphics[width=.74\linewidth]{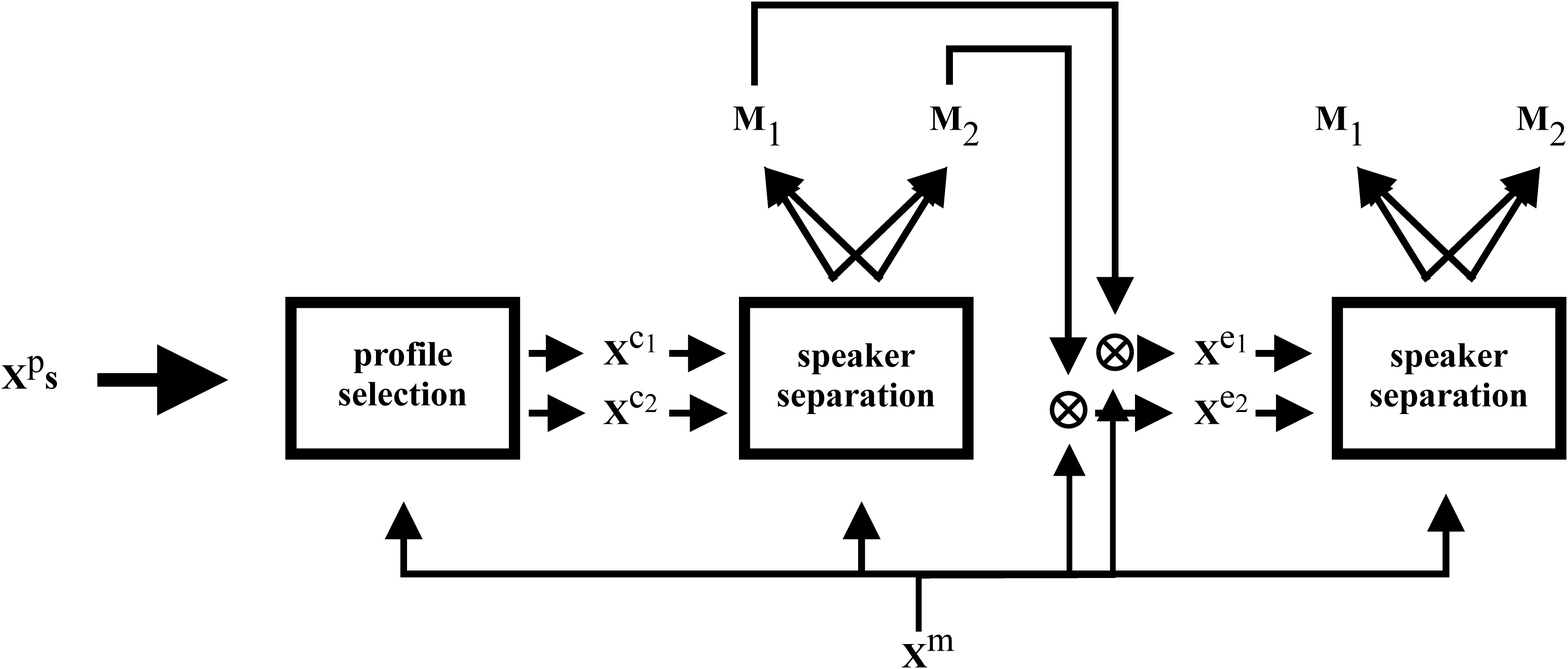}}
\caption{Illustration of SSUES-JT.}
\label{fig:ssues-jt}
\end{figure*}

\subsection{SSUSI-SEP, SSUSI-PSE and SSUSI-JT}
\label{ssec:ssusi_pse_jt}
There are three modules in SSUSI that contain learnable parameters, i.e. the embedding module in the profile selection stage, the embedding module in the speaker separation stage, and the separator in the speaker separation stage. We can thus design three methods to train SSUSI, namely SSUSI that only trains the speaker separation stage (SSUSI-SEP), SSUSI with profile selection embedding (SSUSI-PSE), and SSUSI with joint training (SSUSI-JT).

SSUSI-SEP only trains the speaker separation stage and re-uses the embedding module for profile selection. The rationale behind this method is that both embedding modules are used for the subsequent correlation calculation, as given in equations (\ref{eq:d}) and (\ref{eq:s_dot}). To train the speaker separation stage, oracle relevant profiles are used as $\mathbf{X}^{c_1}$ and $\mathbf{X}^{c_2}$. By sharing the embedding module, the size of the whole model is also reduced.

SSUSI-PSE encourages the correct selection of relevant profiles. In addition to the speaker separation stage, we train the embedding module in the profile selection stage using a specifically designed training objective. The speaker separation stage and the profile selection stage are trained separately. The loss function to train the embedding module in the profile selection stage is as follows:
\begin{equation}
    \label{eq:selection}
    L(\theta) = (1 - w^{o_1} - w^{o_2})^2 + \sum_{o_k \in \mathbf{P}-\{o_1, o_2\}} (w^{o_k})^2
\end{equation}
where $o_1$ and $o_2$ are the oracle relevant profiles, and $w^{o_1}$ and $w^{o_2}$ are the corresponding correlation values calculated by equation (\ref{eq:average}). Speaker inventory $\mathbf{P}$ is divided into two subsets, oracle relevant profiles $\{o_1, o_2\}$ and irrelevant profiles $\mathbf{P} - \{o_1, o_2\}$. For relevant profiles, the training objective is to make their summation equal one, whereas for each irrelevant profile $o_k \in \mathbf{P} - \{o_1, o_2\}$, the objective is to set its weight to zero.

Different from SSUSI-SEP and SSUSI-PSE, which train the speaker separation stage and the profile selection stage separately, SSUSI-JT jointly optimizes the whole SSUSI framework using a single PIT objective. This way, the speaker separation stage may be more robust to wrong profile selections of the profile selection stage. Note that there is an argmax function in the profile selection stage, as shown in equations (\ref{eq:argtop1}) and (\ref{eq:argtop2}). During back-propagation, although the gradients with respect to the indices selected by argmax are hard to derive, we can still calculate the gradients with respect to the selected profiles. Fig. \ref{fig:ssusi_jt} shows a diagram illustrating SSUSI-JT.


\section{Speaker Separation Using Estimated Speech}
\label{sec:ssues}
Although SSUSI can substantially improve separation performance and efficacy, it has two limitations. First, when the number of candidate speakers is large, the profile selection stage in SSUSI tends to select a wrong profile. Second, SSUSI may downgrade to a simple PIT-based separation stage when speaker inventories are not available. The separation performance in such cases would be relatively low.

SSUES solves SSUSI's problems by treating estimated speech (i.e. speaker separation output) $\mathbf{X}^{e_1}$ and $\mathbf{X}^{e_2}$ as speaker profiles. SSUES can be performed iteratively by feeding the estimated speech from a previous iteration to a subsequent separation iteration. Since estimated speech is part of overlapped speech, it is guaranteed to be from a relevant speaker. The negative influence of wrong profile selection can thus be alleviated. Moreover, SSUES provides a feedback loop for both SSUSI and speaker separation, which is able to improve separation performance after each iteration. An illustration of SSUES is presented in Fig. \ref{fig:ssues}.


\subsection{SSUES}
\label{ref:ssec:ssues}
SSUES requires first-pass separation to get initial estimated speech. As mentioned above, the first-pass separation can be SSUSI when a speaker inventory is available, or a speaker separation approach such as PIT when there is no speaker inventory.

After obtaining initial estimated masks $\mathbf{M}_1$ and $\mathbf{M}_2$, we calculate estimated speech as:
\begin{equation}
    \label{eq:iter_1}
    \mathbf{X}^{e_k} = \mathbf{M}_k \otimes \mathbf{X}^{m}
\end{equation}
where $e_k$ indicates estimated speech and $k\in\{1,2\}$ is the speaker index.

Note that for notational simplicity, we only present the spectral magnitude representation of estimated speech in equation (\ref{eq:iter_1}). In implementation, the input to SSUES may be other types of feature and equation (\ref{eq:iter_1}) may change accordingly.

\subsection{SSUES-NT and SSUES-JT}
\label{ssec:ssues-jt}
SSUES can be performed by re-using the speaker separation stage in SSUSI. We denote this no-training method as SSUES-NT. This method, however, may cause an input data mismatch problem between training and test. During training, speaker profiles can be viewed as ``clean'' speech, whereas at test time, estimated speech may contain distortions. To handle this problem, we design SSUES with joint training (SSUES-JT), which jointly optimizes SSUSI and SSUES by a single PIT objective on the output of SSUES. Fig. \ref{fig:ssues-jt} depicts SSUES-JT.


\subsection{Speaker Separation Using Speaker Inventory and Estimated Speech}
\label{ssec:ssusies}
SSUSI and SSUES are closely related. The speaker separation stage in SSUSI makes it possible for SSUES to use estimated speech, and SSUES expands the application scenarios of SSUSI to cases when the number of candidate speakers is large or a speaker inventory is missing. We thus integrate SSUSI and SSUES into SSUSIES. The key component in SSUSIES is the speaker separation stage that leverages bias information such as speaker profiles and estimated speech. When a speaker inventory is available, SSUSIES performs SSUSI for first-pass separation and uses SSUES to leverage the information in estimated speech. In speaker separation tasks, conventional speaker separation is performed for first-pass separation and the SSUES method in SSUSIES can be used to further improve the separation result.

\section{Experimental Setup}
\label{sec:exp}
\subsection{Dataset}
\label{ssec:exp_dataset}
Our experiments are conducted on the LibriSpeech corpus \cite{Panayotov2015Librispeech:Books} following the same recipe as \cite{xiao2019single}. The training set is generated using both train-clean-100 and train-clean-360. At test time, overlapped speech is generated using the test-clean set. There are 1172 speakers in the training set and 40 other speakers in the test set.

We use globally mean-variance normalized log spectral magnitude features as input. The length of each frame is 32 ms (i.e. 512 samples with a sampling rate of 16 kHz) and the shift between frames is 16 ms. The waveform signals are transformed using 512 dimensional short-time Fourier transformation. For training targets $M$s, we use the spectral magnitude mask \cite{wang2014training}.

\subsection{Baseline Systems}
\label{ssec:exp_baseline}
There are two baseline systems in this study, an utterance-wise PIT-based speaker separation model \cite{Kolbk2017SpeechSystems} and Xiao \emph{et al.}'s speech extraction system \cite{xiao2019single}. The PIT-based model consists of six bidirectional long short-term memory (BLSTM) layers, each of which has 512 nodes. The speech extraction system has the same number of learnable parameters as the PIT-based model. For both the PIT-based model and speech extraction system, the optimizer is Adam and the learning rate is $10^{-4}$.

\subsection{SSUSI}
\label{ssec:exp_ssusi}
As mentioned in Section \ref{ssec:ssusi_pse_jt}, there are three learnable modules in SSUSI, i.e. two embedding modules and the separator in the speaker separation stage. The two embedding modules have the same architecture, which consists of three BLSTM layers. The separator also has three BLSTM layers. All the BLSTM layers contain 512 nodes. SSUSI-SEP and SSUSI-JT have the same number of learnable parameters as those in the baselines, whereas SSUSI-PSE has an additional three-layer embedding module for profile selection.

The  SSUSI-SEP is trained using oracle relevant profiles. For SSUSI-PSE and SSUSI-JT, four speaker profiles, two relevant and two irrelevant, are used as the speaker inventory during training. The embedding module in the profile selection stage of SSUSI-PSE is initialized with the well-trained embedding module in SSUSI-SEP. For SSUSI-JT, the whole model is initialized with SSUSI-SEP. At the speaker separation stage, speaker biases are concatenated with the embeddings and the original features of overlapped speech along the feature dimension. The learning rate for SSUSI-SEP, SSUSI-PSE, and SSUSI-JT are $10^{-4}$, $10^{-6}$, and $10^{-5}$, respectively. All the other hyper-parameters are the same as the baselines. To avoid over-fitting, we apply online simulation, which generates the overlapped speech during model training. The model checkpoint used for evaluation is thus selected based on training loss.

Note that during training, we shuffle the order of speaker profiles. This makes the separation performance of SSUSI uninfluenced by the order of speaker profiles in the speaker inventory.

\subsection{SSUES}
\label{ssec:exp_ssues}
SSUES has the same number of learnable parameters as the separator in SSUSI. It contains 3 BLSTM layers, each of which consists of 512 nodes.

As mentioned in Section \ref{ssec:exp_dataset}, we use globally normalized log spectrum magnitude features in this study. Therefore, during SSUES-JT training, we perform logarithm and mean-variance normalization in addition to the spectral magnitude masking shown in equation (\ref{eq:iter_1}).

\subsection{ASR Backend}
\label{ssec:exp_asr}
Our ASR backend is a DNN-HMM hybrid model trained on the clean training set of LibriSpeech. The model has three BLSTM layers, each of which contains 512 nodes. We generate forced aligned senone labels using Kaldi \cite{Povey2011TheToolkit} and train the model under the maximum mutual information (MMI) criterion using PyTorch. The word error rate (WER) of this model on non-overlapped LibriSpeech test set is 5.7\%.

\section{Evaluation Results}
\label{sec:result}
We first present the results of SSUSI and then show how SSUES improves both PIT and SSUSI.

\subsection{SSUSI}
\label{ssec:re_sussi}
\begin{table}[ht]
    \centering
    \begin{tabular}{c c c c c}
        \toprule
        method & \# ir-profiles & $\geq$1 (\%) & 2 (\%) & SDR (dB) \\
        \midrule
        \multirow{7}{*}{\shortstack{SSUSI-SEP}} & 0 & 100 & 100 & 12.1 \\
                                                         & 1 & 100 & 82.1 & 11.8 \\
                                                         & 2 & 99.9 & 71.6 & 11.6 \\
                                                         & 3 & 99.8 & 64.1 & 11.4 \\
                                                         & 4 & 99.5 & 58.2 & 11.2 \\
                                                         & 5 & 99.2 & 54.9 & 11.1 \\
                                                         & 6 & 99.0 & 51.4 & 11.0 \\
        \midrule
        \multirow{7}{*}{\shortstack{SSUSI-PSE}} & 0 & 100 & 100 & 12.1 \\
                                                         & 1 & 100 & 86.7 & 11.9 \\
                                                         & 2 & 100 & 78.5 & 11.7 \\
                                                         & 3 & 99.8 & 72.5 & 11.6 \\
                                                         & 4 & 99.7 & 67.8 & 11.5 \\
                                                         & 5 & 99.4 & 63.8 & 11.3 \\
                                                         & 6 & 99.3 & 61.1 & 11.3 \\
        \midrule
        \multirow{7}{*}{\shortstack{SSUSI-JT}} & 0 & 100 & 100 & 12.2 \\
                                                         & 1 & 100 & 81.0 & 12.0 \\
                                                         & 2 & 99.8 & 69.6 & 11.9 \\
                                                         & 3 & 99.6 & 61.9 & 11.8 \\
                                                         & 4 & 99.4 & 56.5 & 11.6 \\
                                                         & 5 & 99.0 & 52.8 & 11.6 \\
                                                         & 6 & 98.7 & 49.7 & 11.5 \\
        \bottomrule
    \end{tabular}
    \caption{SDRs and correct profile selection rates of SSUSI-SEP, SSUSI-PSE, and SSUSI-JT. The number of profiles corresponding to irrelevant speakers is denoted as \emph{\# ir-profiles}. The total number of profiles in the speaker inventory is \emph{\# ir-profiles} plus 2. The correct selection of at least one relevant profile is denoted as \emph{$\geq$1} and the correct selection of both relevant profiles is \emph{2}.}
    \label{tab:ssusi}
\end{table}

Table \ref{tab:ssusi} contains the signal to distortion ratios (SDRs) of the three training approaches, SSUSI-SEP, SSUSI-PSE, and SSUSI-JT. We also list correct profile selection rates, which measure the performance of the profile selection stage. The SDR of unprocessed mixtures is 0.0 dB. SSUSI-SEP gets an SDR of 12.1 dB when both relevant profiles are correctly selected. With the increase of irrelevant profiles, all three metrics decrease. From 0 to 6 irrelevant profiles, the correct profile selection rate of at lease one relevant profile drops slightly from 100\% to 99.0\%, whereas the correct selection rate of both relevant profiles decreases significantly from 100\% to 51.4\%. SDRs are degraded by wrong profile selections. The SDR on 6 irrelevant profiles drops to 11.0 dB. SSUSI-PSE improves correct profile selection rates substantially by using the additional profile selection embedding module. The improvement gets larger as the number of irrelevant profiles increases. SDRs benefit from better profile selection. Compared with that of SSUSI-SEP, the SDR of SSUSI-PSE on 6 irrelevant profiles increases by 0.3 dB. SSUSI-JT is able to achieve substantial SDR improvement over SSUSI-SEP without increasing the model size. Its SDR with 6 irrelevant profiles is 11.5 dB, outperforming both SSUSI-SEP and SSUSI-PSE. The SDR of SSUSI-JT on 0 irrelevant profile is slightly better than those of SSUSI-SEP and SSUSI-PSE. Note that the correct profile selection rates of SSUSI-JT are worse than those of SSUSI-PSE and even those of SSUSI-SEP. This shows that SSUSI-JT is robust to wrong profile selections. Since SSUSI-JT yields the best SDR results without using extra learnable parameters, we denote it as SSUSI in the remainder of this paper.

\begin{table}[ht]
    \centering
    \begin{tabular}{c c c c}
        \toprule
        method & \# ir-profiles & SDR (dB) & WER (\%) \\
        \midrule
        PIT & - & 8.7 & 36.5 \\
        \midrule
        \multirow{4}{*}{SSUSI} & 0 & 12.2 & 19.1 \\
                                                         & 6 & 11.5 & 21.8 \\
                                                         & 22 & 11.0 & 23.4 \\
                                                         & 30 & 10.8 & 24.1 \\
        \bottomrule
    \end{tabular}
    \caption{SDR and WER comparisons between SSUSI and PIT. See Table \ref{tab:ssusi} caption for acronyms.}
    \label{tab:pit_ssusi}
\end{table}

Table \ref{tab:pit_ssusi} shows the SDR and WER comparisons between SSUSI and PIT. Because of the ability to leverage speaker information, SSUSI performs significantly better than PIT in both SDR and WER. In the case of 30 irrelevant profiles (i.e. 32 candidate profiles in the speaker inventory), SSUSI still yields an SDR of 10.8 dB, which is substantially better than the 8.7 dB SDR of PIT. Note that SSUSI is trained using only 2 irrelevant profiles. The results with 6, 22, and 30 irrelevant profiles demonstrate the robustness of SSUSI. In terms of WERs, SSUSI outperforms PIT by 48\% relatively in the case of 0 irrelevant profile. When there are 30 irrelevant profiles, the relative improvement is still 34\%. Note that all the WERs in Table \ref{tab:pit_ssusi} are relatively high for the LibriSpeech corpus. This is due to the distortions in estimated speech.


\begin{table}[ht]
    \centering
    \begin{tabular}{c c c c}
        \toprule
        method & \# ir-profiles & SDR (dB) & WER (\%) \\
        \midrule
        \multirow{3}{*}{Speech Extraction \cite{xiao2019single}} & 0 & 11.5 & 21.9 \\
                                                         & 1 & 11.1 & 23.3 \\
                                                         & 2 & 10.9 & 24.4 \\
        \midrule
        \multirow{3}{*}{SSUSI} & 0 & 12.2 & 19.1 \\
                                                         & 1 & 12.0 & 19.9 \\
                                                         & 2 & 11.9 & 20.4 \\
        \bottomrule
    \end{tabular}
    \caption{SDR and WER comparisons between SSUSI and a speech extraction system. See Table \ref{tab:ssusi} caption for acronyms.}
    \label{tab:extraction_ssusi}
\end{table}

Table \ref{tab:extraction_ssusi} presents the SDR and WER comparisons between SSUSI and the speech extraction system \cite{xiao2019single}. SSUSI substantially outperforms the speech extraction baseline. In terms of SDR, an improvement of more than 0.7 dB is yielded. For WER, the overall relative improvement is over 13\%. This suggests that SSUSI is better at discriminating speaker pairs by the awareness of a competing speaker. In addition to the improvement in separation performance, SSUSI is significantly more efficient than the speech extraction system. In the case of 0 irrelevant profile, the computation time reduction during test is about 40\% relatively. When there are 30 irrelevant profiles, the computation time reduction is about 70\% relatively. The reason of this efficiency improvement is that speech extraction needs to launch one model instance for each candidate speaker, whereas SSUSI filters out all but one pair of speaker profiles for speaker separation.

\begin{table}[ht]
    \centering
    \begin{tabular}{c c c c c}
        \toprule
        method & \# ir-profiles & standard & m1 & m2 \\
        \midrule
        \multirow{7}{*}{\shortstack{SSUSI}} & 0 & 12.2 & - & - \\
                                                         & 1 & 12.0 & 10.0 & - \\
                                                         & 2 & 11.9 & 10.2 & 8.6 \\
                                                         & 3 & 11.8 & 10.2 & 8.6 \\
                                                         & 4 & 11.6 & 10.1 & 8.5 \\
                                                         & 5 & 11.6 & 10.1 & 8.5 \\
                                                         & 6 & 11.5 & 10.1 & 8.3 \\
        \bottomrule
    \end{tabular}
    \caption{SDRs of SSUSI in cases when one or both relevant profiles are not in the speaker inventory. \emph{standard} denotes both relevant profiles are in the speaker inventory, \emph{m1} refers to the case when one relevant profile is missing, and \emph{m2} means both relevant profiles are missing.}
    \label{tab:analysis_one}
\end{table}

Table \ref{tab:analysis_one} provides the SDRs of SSUSI in cases when one or both relevant profiles are missing from the speaker inventory. This corresponds to the real-world scenario when one or more unregistered speakers attend the meeting and some of them are involved in overlapped speech. In the case when one relevant profile is missing, the SDRs of SSUSI drop to values close to 10.1 dB, which is still substantially better than the 8.7 dB result of PIT. This suggests that SSUSI is able to leverage the information in the remaining relevant profile even when the other one is missing. When both relevant profiles are missing, the SDRs of SSUSI are around 8.5 dB, which is similar to the SDR of PIT. This indicates that when both speakers in the overlapped speech are unregistered speakers, SSUSI performs similarly to PIT. Note that speech extraction cannot work in above cases when there are unregistered speakers.

\subsection{SSUES}
\label{ssec:ssues}
\begin{table}[ht]
    \centering
    \begin{tabular}{c c c c c c}
        \toprule
        method & \# ir-profiles & no-iter & iter1 & iter2 & iter3 \\
        \midrule
        \multirow{9}{*}{\shortstack{SSUES-NT}} & 0 & 12.2 & 12.4 & 12.4 & 12.4 \\
                                                         & 1 & 12.0 & 12.3 & 12.3 & 12.3 \\
                                                         & 2 & 11.9 & 12.2 & 12.2 & 12.2 \\
                                                         & 3 & 11.8 & 12.1 & 12.2 & 12.2 \\
                                                         & 4 & 11.6 & 12.0 & 12.1 & 12.1 \\
                                                         & 5 & 11.6 & 12.0 & 12.1 & 12.1 \\
                                                         & 6 & 11.5 & 11.9 & 12.0 & 12.0 \\
                                                         & 22 & 11.0 & 11.5 & 11.7 & 11.7 \\
                                                         & 30 & 10.8 & 11.4 & 11.6 & 11.7 \\
        \midrule
        \multirow{9}{*}{\shortstack{SSUES-JT}} & 0 & 12.2 & 12.3 & 12.3 & 12.4 \\
                                                         & 1 & 12.0 & 12.2 & 12.3 & 12.3 \\
                                                         & 2 & 11.9 & 12.1 & 12.2 & 12.3 \\
                                                         & 3 & 11.8 & 12.1 & 12.2 & 12.2 \\
                                                         & 4 & 11.6 & 12.0 & 12.1 & 12.2 \\
                                                         & 5 & 11.6 & 12.0 & 12.1 & 12.1 \\
                                                         & 6 & 11.5 & 11.9 & 12.0 & 12.1 \\
                                                         & 22 & 11.0 & 11.6 & 11.8 & 11.8 \\
                                                         & 30 & 10.8 & 11.5 & 11.7 & 11.8 \\
        \bottomrule
    \end{tabular}
    \caption{SDRs of SSUES-NT and SSUES-JT using SSUSI as first-pass separation. \emph{no-iter} refers to first-pass separation. \emph{iter1}, \emph{iter2}, and \emph{iter3} denote the first, second, and third SSUES based separation iteration, respectively.}
    \label{tab:ssues}
\end{table}

Table \ref{tab:ssues} shows the comparison between SSUES-NT and SSUES-JT. Both of them use SSUSI as first-pass separation. For SSUES-NT, the improvement over SSUSI is substantial, especially when the number of irrelevant profiles is large. With 30 irrelevant profiles, the improvement after three iterations is about 1 dB. More specifically, the SDR of SSUES-NT with 30 irrelevant profiles is comparable to that of SSUSI with 3 irrelevant profiles. These results show that SSUES-NT is able to alleviate the wrong profile selection problem of SSUSI and consistently improve the separation performance. Note that the performance of SSUES-NT with 0 irrelevant profile is also better than that of SSUSI. The reason is that, in addition to speaker information, SSUES-NT can leverage the contextual information in estimated speech. SSUES-NT performs similarly to SSUES-JT even without joint training. This shows that trained with a large number of speaker profiles, SSUES-NT is able to generalize to estimated speech. Considering the fact that SSUES-JT is better than SSUES-NT by at most 0.1 dB and that SSUES is designed to work with various first-pass separation approaches, we use SSUES-NT as SSUES in the remainder of this paper. 

\begin{table}[ht]
    \centering
    \begin{tabular}{c c c c}
        \toprule
        method & \# iter & SDR (dB) & WER (\%) \\
        \midrule
        PIT & - & 8.7 & 36.5 \\
        \midrule
        \multirow{3}{*}{\shortstack{ + SSUES}} & 1 & 10.5 & 24.8 \\
        & 2 & 10.8 & 23.2 \\
        & 3 & 10.9 & 22.9 \\
        \bottomrule
    \end{tabular}
    \caption{SDR and WER comparisons between PIT and PIT + SSUES. \emph{\# iter} denotes the number of SSUES based separation iterations.}
    \label{tab:ssues_pit}
\end{table}

Table \ref{tab:ssues_pit} presents the SDR and WER comparisons between PIT and PIT + SSUES. With only one iteration of SSUES, the SDR improvement is already 1.8 dB and the WER improvement is 31\% relatively. After three iterations, the WER improvement is increased to 37\% relatively. These comparisons clearly show the efficacy of SSUES in improving the performance of PIT.

\begin{table}[ht]
    \centering
    \begin{tabular}{c c c c}
        \toprule
        method & \# iter & SDR (dB) & WER (\%) \\
        \midrule
        SSUSI & - & 10.8 & 24.1 \\
        \midrule
        \multirow{3}{*}{\shortstack{ + SSUES}} & 1 & 11.4 & 21.1 \\
        & 2 & 11.6 & 20.4 \\
        & 3 & 11.7 & 20.3 \\
        \bottomrule
    \end{tabular}
    \caption{SDR and WER comparisons between SSUSI and SSUSI + SSUES with 30 irrelevant profiles.}
    \label{tab:ssues_ssusi}
\end{table}

Table \ref{tab:ssues_ssusi} shows the SDR and WER comparisons between SSUSI and SSUSI + SSUES with 30 irrelevant profiles. With three iterations of SSUES, the SDR improvement is 0.9 dB and the WER reduction is 16\% relatively. Note that the SDR and WER results of SSUSI with 2 irrelevant profiles are 11.9 dB and 20.4\%, as shown in Table \ref{tab:extraction_ssusi}. The similar results of SSUSI with 2 irrelevant profiles and SSUSI + SSUES with 30 irrelevant profiles show that SSUES can substantially improve the performance of SSUSI.

\section{Concluding Remarks}
\label{sec:conclusion}
We have proposed SSUSIES, a speaker separation framework that is capable of leveraging external information such as speaker profiles and estimated speech. Compared with speech extraction, SSUSIES achieves more than 13\% relative improvement in WER and up to 70\% relative improvement in computational efficiency. In addition, SSUSIES outperforms PIT by 13\% relatively in WER. Future research will extend SSUSIES to multi-channel conditions and evaluate SSUSIES in real conversations.

\bibliographystyle{IEEEtranS}
\bibliography{references,bibtex}

\begin{thebibliography}{10}
\providecommand{\url}[1]{#1}
\csname url@samestyle\endcsname
\providecommand{\newblock}{\relax}
\providecommand{\bibinfo}[2]{#2}
\providecommand{\BIBentrySTDinterwordspacing}{\spaceskip=0pt\relax}
\providecommand{\BIBentryALTinterwordstretchfactor}{4}
\providecommand{\BIBentryALTinterwordspacing}{\spaceskip=\fontdimen2\font plus
\BIBentryALTinterwordstretchfactor\fontdimen3\font minus
  \fontdimen4\font\relax}
\providecommand{\BIBforeignlanguage}[2]{{%
\expandafter\ifx\csname l@#1\endcsname\relax
\typeout{** WARNING: IEEEtranS.bst: No hyphenation pattern has been}%
\typeout{** loaded for the language `#1'. Using the pattern for}%
\typeout{** the default language instead.}%
\else
\language=\csname l@#1\endcsname
\fi
#2}}
\providecommand{\BIBdecl}{\relax}
\BIBdecl

\bibitem{chen2018sequence}
Z.~Chen and J.~Droppo, ``Sequence modeling in unsupervised single-channel
  overlapped speech recognition,'' in \emph{Proc. of International Conference
  on Acoustics, Speech and Signal Processing (ICASSP)}.\hskip 1em plus 0.5em
  minus 0.4em\relax IEEE, 2018, pp. 4809--4813.

\bibitem{chen2017deep}
Z.~Chen, Y.~Luo, and N.~Mesgarani, ``Deep attractor network for
  single-microphone speaker separation,'' in \emph{Proc. of International
  Conference on Acoustics, Speech and Signal Processing (ICASSP)}.\hskip 1em
  plus 0.5em minus 0.4em\relax IEEE, 2017, pp. 246--250.

\bibitem{chen2018location}
Z.~Chen, X.~Xiao, T.~Yoshioka, J.~Li, H.~Erdogan, and Y.~Gong, ``Multi-channel
  multi-speaker overlapped speech recognition with location guided speech
  extraction network,'' in \emph{Spoken Language Technology Workshop (SLT)},
  2018, pp. 558--565.

\bibitem{delcroix2018single}
M.~Delcroix, K.~Zmolikova, K.~Kinoshita, A.~Ogawa, and T.~Nakatani, ``Single
  channel target speaker extraction and recognition with speaker beam,'' in
  \emph{Proc. of International Conference on Acoustics, Speech and Signal
  Processing (ICASSP)}.\hskip 1em plus 0.5em minus 0.4em\relax IEEE, 2018, pp.
  5554--5558.

\bibitem{ephrat2018looking}
A.~Ephrat, I.~Mosseri, O.~Lang, T.~Dekel, K.~Wilson, A.~Hassidim, W.~T.
  Freeman, and M.~Rubinstein, ``Looking to listen at the cocktail party: A
  speaker-independent audio-visual model for speech separation,'' \emph{arXiv
  preprint arXiv:1804.03619}, 2018.

\bibitem{erdogan2015phase}
H.~Erdogan, J.~R. Hershey, S.~Watanabe, and J.~Le~Roux, ``Phase-sensitive and
  recognition-boosted speech separation using deep recurrent neural networks,''
  in \emph{Proc. of IEEE International Conference on Acoustics, Speech and
  Signal Processing (ICASSP)}.\hskip 1em plus 0.5em minus 0.4em\relax IEEE,
  2015, pp. 708--712.

\bibitem{gu2019end}
R.~Gu, J.~Wu, S.~X. Zhang, L.~Chen, Y.~Xu, M.~Yu, D.~Su, Y.~Zou, and D.~Yu,
  ``End-to-end multi-channel speech separation,'' \emph{arXiv preprint
  arXiv:1905.06286}, 2019.

\bibitem{hershey2016deep}
J.~R. Hershey, Z.~Chen, J.~Le~Roux, and S.~Watanabe, ``Deep clustering:
  Discriminative embeddings for segmentation and separation,'' in \emph{Proc.
  of International Conference on Acoustics, Speech and Signal Processing
  (ICASSP)}.\hskip 1em plus 0.5em minus 0.4em\relax IEEE, 2016, pp. 31--35.

\bibitem{Hu2013AnSeparation}
K.~Hu and D.~L. Wang, ``{An iterative model-based approach to cochannel speech
  separation},'' \emph{EURASIP Journal on Audio, Speech, and Music Processing},
  2013.

\bibitem{isik2016single}
Y.~Isik, J.~Le~Roux, Z.~Chen, S.~Watanabe, and J.~R. Hershey, ``Single-channel
  multi-speaker separation using deep clustering,'' in \emph{Proc. of
  INTERSPEECH}, 2016, pp. 545--549.

\bibitem{kolbaek2017multitalker}
M.~Kolb{\ae}k, D.~Yu, Z.~H. Tan, and J.~Jensen, ``Multitalker speech separation
  with utterance-level permutation invariant training of deep recurrent neural
  networks,'' \emph{IEEE/ACM Transactions on Audio, Speech and Language
  Processing (TASLP)}, no.~10, pp. 1901--1913, 2017.

\bibitem{Kolbk2017SpeechSystems}
M.~Kolbk, Z.~H. Tan, and J.~Jensen, ``{Speech intelligibility potential of
  general and specialized deep neural network based speech enhancement
  systems},'' \emph{IEEE/ACM Transactions on Audio, Speech and Language
  Processing (TASLP)}, pp. 153--167, 2017.

\bibitem{li2018deep}
L.~Li and H.~Kameoka, ``Deep clustering with gated convolutional networks,'' in
  \emph{Proc. of International Conference on Acoustics, Speech and Signal
  Processing (ICASSP)}.\hskip 1em plus 0.5em minus 0.4em\relax IEEE, 2018, pp.
  16--20.

\bibitem{li2018source}
Z.~X. Li, Y.~Song, L.~R. Dai, and I.~McLoughlin, ``Source-aware context network
  for single-channel multi-speaker speech separation,'' in \emph{Proc. of
  International Conference on Acoustics, Speech and Signal Processing
  (ICASSP)}.\hskip 1em plus 0.5em minus 0.4em\relax IEEE, 2018, pp. 681--685.

\bibitem{li2019listening}
------, ``Listening and grouping: an online autoregressive approach for
  monaural speech separation,'' \emph{IEEE/ACM Transactions on Audio, Speech
  and Language Processing (TASLP)}, vol.~27, no.~4, pp. 692--703, 2019.

\bibitem{Liu2019DivideSeparation}
Y.~Liu and D.~L. Wang, ``{Divide and Conquer: A Deep CASA Approach to
  Talker-Independent Monaural Speaker Separation},'' \emph{IEEE/ACM
  Transactions on Audio Speech and Language Processing (TASLP)}, vol.~27, pp.
  2092--2102, 2019.

\bibitem{luo2018speaker}
Y.~Luo, Z.~Chen, and N.~Mesgarani, ``Speaker-independent speech separation with
  deep attractor network,'' \emph{IEEE/ACM Transactions on Audio, Speech, and
  Language Processing (TASLP)}, vol.~26, no.~4, pp. 787--796, 2018.

\bibitem{luo2019augmented}
Y.~Luo and N.~Mesgarani, ``Augmented time-frequency mask estimation in
  cluster-based source separation algorithms,'' in \emph{Proc. of International
  Conference on Acoustics, Speech and Signal Processing (ICASSP)}.\hskip 1em
  plus 0.5em minus 0.4em\relax IEEE, 2019, pp. 710--714.

\bibitem{luo2019conv}
------, ``Conv-{TasNet}: Surpassing ideal time--frequency magnitude masking for
  speech separation,'' \emph{IEEE/ACM Transactions on Audio, Speech, and
  Language Processing (TASLP)}, vol.~27, pp. 1256--1266, 2019.

\bibitem{ochiai2019unified}
T.~Ochiai, M.~Delcroix, K.~Kinoshita, A.~Ogawa, and T.~Nakatani, ``A unified
  framework for neural speech separation and extraction,'' in \emph{Proc. of
  International Conference on Acoustics, Speech and Signal Processing
  (ICASSP)}.\hskip 1em plus 0.5em minus 0.4em\relax IEEE, 2019, pp. 6975--6979.

\bibitem{Panayotov2015Librispeech:Books}
V.~Panayotov, G.~Chen, D.~Povey, and S.~Khudanpur, ``{Librispeech: An ASR
  corpus based on public domain audio books},'' in \emph{Proc. of International
  Conference on Acoustics, Speech and Signal Processing (ICASSP)}, 2015, pp.
  5206--5210.

\bibitem{perotin2018multichannel}
L.~Perotin, R.~Serizel, E.~Vincent, and A.~Gu{\'e}rin, ``Multichannel speech
  separation with recurrent neural networks from high-order ambisonics
  recordings,'' in \emph{Proc. of International Conference on Acoustics, Speech
  and Signal Processing (ICASSP)}, 2018, pp. 36--40.

\bibitem{Povey2011TheToolkit}
D.~Povey, A.~Ghoshal, G.~Boulianne, L.~Burget, O.~Glembek, N.~Goel,
  M.~Hannemann, P.~Motlicek, Y.~Qian, P.~Schwarz, and {others}, ``{The Kaldi
  speech recognition toolkit},'' in \emph{Proc. of IEEE Workshop on Automatic
  Speech Recognition and Understanding (ASRU)}, 2011, pp. 1--4.

\bibitem{shi2019furcanext}
Z.~Shi, H.~Lin, L.~Liu, R.~Liu, and J.~Han, ``Furcanext: End-to-end monaural
  speech separation with dynamic gated dilated temporal convolutional
  networks,'' \emph{arXiv preprint arXiv:1902.04891}, 2019.

\bibitem{Wang2018DeepMixtures}
J.~Wang, J.~Chen, D.~Su, L.~Chen, M.~Yu, Y.~Qian, and D.~Yu, ``{Deep extractor
  network for target speaker recovery from single channel speech mixtures},''
  in \emph{Proc. of INTERSPEECH}, 2018, pp. 307--311.

\bibitem{Wang2019SpeechInventory}
P.~Wang, Z.~Chen, X.~Xiao, Z.~Meng, T.~Yoshioka, T.~Zhou, L.~Lu, and J.~Li,
  ``{Speech Separation Using Speaker Inventory},'' in \emph{Proc. of IEEE
  Workshop on Automatic Speech Recognition and Understanding (ASRU)}, 2019, pp.
  230--236.

\bibitem{wang2018voicefilter}
Q.~Wang, H.~Muckenhirn, K.~Wilson, P.~Sridhar, Z.~Wu, J.~R. Hershey, R.~A.
  Saurous, R.~J. Weiss, Y.~Jia, and I.~L. Moreno, ``Voicefilter: Targeted voice
  separation by speaker-conditioned spectrogram masking,'' \emph{arXiv preprint
  arXiv:1810.04826}, 2018.

\bibitem{wang2014training}
Y.~Wang, A.~Narayanan, and D.~Wang, ``On training targets for supervised speech
  separation,'' \emph{IEEE/ACM Transactions on Audio, Speech and Language
  Processing (TASLP)}, vol.~22, pp. 1849--1858, 2014.

\bibitem{wang2018alternative}
Z.~Q. Wang, J.~Le~Roux, and J.~R. Hershey, ``Alternative objective functions
  for deep clustering,'' in \emph{Proc. of International Conference on
  Acoustics, Speech and Signal Processing (ICASSP)}.\hskip 1em plus 0.5em minus
  0.4em\relax IEEE, 2018, pp. 686--690.

\bibitem{wang2018multi}
------, ``Multi-channel deep clustering: Discriminative spectral and spatial
  embeddings for speaker-independent speech separation,'' in \emph{Proc. of
  International Conference on Acoustics, Speech and Signal Processing
  (ICASSP)}, 2018, pp. 1--5.

\bibitem{weninger2014discriminatively}
F.~Weninger, J.~R. Hershey, J.~Le~Roux, and B.~Schuller, ``Discriminatively
  trained recurrent neural networks for single-channel speech separation,'' in
  \emph{Proc. of IEEE Global Conference on Signal and Information Processing
  (GlobalSIP)}.\hskip 1em plus 0.5em minus 0.4em\relax IEEE, 2014, pp.
  577--581.

\bibitem{wu2019time}
J.~Wu, Y.~Xu, S.~X. Zhang, L.~W. Chen, M.~Yu, L.~Xie, and D.~Yu, ``Time domain
  audio visual speech separation,'' \emph{arXiv preprint arXiv:1904.03760},
  2019.

\bibitem{xiao2019single}
X.~Xiao, Z.~Chen, T.~Yoshioka, H.~Erdogan, C.~Liu, D.~Dimitriadis, J.~Droppo,
  and Y.~Gong, ``Single-channel speech extraction using speaker inventory and
  attention network,'' in \emph{Proc. of International Conference on Acoustics,
  Speech and Signal Processing (ICASSP)}.\hskip 1em plus 0.5em minus
  0.4em\relax IEEE, 2019, pp. 86--90.

\bibitem{xu2019optimization}
C.~Xu, W.~Rao, E.~S. Chng, and H.~Li, ``Optimization of speaker extraction
  neural network with magnitude and temporal spectrum approximation loss,'' in
  \emph{Proc. of International Conference on Acoustics, Speech and Signal
  Processing (ICASSP)}.\hskip 1em plus 0.5em minus 0.4em\relax IEEE, 2019, pp.
  6990--6994.

\bibitem{xu2018single}
C.~Xu, W.~Rao, X.~Xiao, E.~S. Chng, and H.~Li, ``Single channel speech
  separation with constrained utterance level permutation invariant training
  using grid lstm,'' in \emph{Proc. of International Conference on Acoustics,
  Speech and Signal Processing (ICASSP)}.\hskip 1em plus 0.5em minus
  0.4em\relax IEEE, 2018, pp. 6--10.

\bibitem{yoshioka2018multi}
T.~Yoshioka, H.~Erdogan, Z.~Chen, and F.~Alleva, ``Multi-microphone neural
  speech separation for far-field multi-talker speech recognition,'' in
  \emph{Proc. of International Conference on Acoustics, Speech and Signal
  Processing (ICASSP)}.\hskip 1em plus 0.5em minus 0.4em\relax IEEE, 2018, pp.
  5739--5743.

\bibitem{yu2017permutation}
D.~Yu, M.~Kolb{\ae}k, Z.~H. Tan, and J.~Jensen, ``Permutation invariant
  training of deep models for speaker-independent multi-talker speech
  separation,'' in \emph{Proc. of International Conference on Acoustics, Speech
  and Signal Processing (ICASSP)}.\hskip 1em plus 0.5em minus 0.4em\relax IEEE,
  2017, pp. 241--245.

\bibitem{zhao2018sound}
H.~Zhao, C.~Gan, A.~Rouditchenko, C.~Vondrick, J.~McDermott, and A.~Torralba,
  ``The sound of pixels,'' \emph{arXiv preprint arXiv:1804.03160}, 2018.

\bibitem{zhao2018two}
Y.~Zhao, Z.~Q. Wang, and D.~L. Wang, ``Two-stage deep learning for
  noisy-reverberant speech enhancement,'' \emph{IEEE/ACM transactions on audio,
  speech, and language processing}, vol.~27, no.~1, pp. 53--62, 2018.

\bibitem{zmolikova2017speaker}
K.~Zmolikova, M.~Delcroix, K.~Kinoshita, T.~Higuchi, A.~Ogawa, and T.~Nakatani,
  ``Speaker-aware neural network based beamformer for speaker extraction in
  speech mixtures,'' in \emph{Proc. of INTERSPEECH}, 2017, pp. 2655--2659.

\end{thebibliography}

\vfill

\end{document}